\documentclass[12pt,preprint]{aastex}

\slugcomment{Accepted by The Astronomical Journal - July 2004}

\shorttitle{Outer Regions of M33}
\shortauthors{Tiede et al.}

\begin{document}
\def\lea{\mathrel{<\kern-1.0em\lower0.9ex\hbox{$\sim$}}}
\def\gea{\mathrel{>\kern-1.0em\lower0.9ex\hbox{$\sim$}}}

\title{The Stellar Populations in the Outer Regions of \\
M33. I. Metallicity Distribution Function\altaffilmark{1}}
    
\author{Glenn P. Tiede\altaffilmark{2}, Ata Sarajedini, and Michael K. Barker}

\affil{Department of Astronomy, University of Florida, P.O. Box 112055, 
Gainesville, FL 32611-2055}

\email{gptiede@bgnet.bgsu.edu, \{ata,mbarker\}@astro.ufl.edu}

\altaffiltext{1}{Based on observations taken with
the WIYN 3.5m telescope. The WIYN Observatory is a joint facility of the 
University of Wisconsin-Madison, Indiana University, Yale University, and 
the National Optical Astronomy Observatories.}

\altaffiltext{2}{Current address: Department of Physics and Astronomy,
104 Overman Hall, Bowling Green State University, Bowling Green, OH 43403}

\begin{abstract}
We present deep CCD photometry in the $VI$ passbands using the WIYN 3.5m
telescope of a field located approximately
20' southeast of the center of M33; this field includes the region studied
by Mould \& Kristian in their 1986 paper. The color-magnitude diagram (CMD) 
extends to I$\sim$25 and shows a prominent red giant branch (RGB),
along with significant numbers of asymptotic giant branch and young
main sequence stars. The red clump of core helium burning stars is also
discernable near the limit of our CMD. The I-band apparent magnitude 
of the red giant branch tip
implies  a distance modulus of $(m-M)_I = 24.77 \pm 0.06$, which combined
with an adopted reddening of $E(V-I)=0.06\pm0.02$ yields an absolute modulus
of $(m-M)_0 = 24.69 \pm 0.07$ (867$\pm$28 kpc) for M33. Over the range
of deprojected radii covered by our field ($\sim$8.5 to $\sim$12.5 kpc),
we find a significant age gradient with an upper limit of 
$\sim$1 Gyr ($\sim$0.25 Gyr/kpc). Comparison of the RGB photometry
to empirical giant branch sequences for Galactic globulars
allows us to use the dereddened color of these
stars to construct a metallicity distribution function (MDF). The primary
peak in the MDF is at a metallicity of $[Fe/H]$$\sim$--1.0 with a tail to
lower abundances. The peak does show radial variation with a slope
of $\Delta$[Fe/H]/$\Delta$$R_{deproj}$ = --0.06 $\pm$ 0.01 dex/kpc.
This gradient is consistent with the variation seen in the inner {\it disk} regions 
of M33. As such, we conclude that the vast majority of stars in this field belong
to the disk of M33, not the halo as previously thought.
\end{abstract}



\section{Introduction}

In recent years M33 has been the target of many systematic studies.  From
searches for variable stars to pin down one of the fundamental
distance estimates in the Cepheid distance latter (e.g. Macri et al. 2001), to
kinematical studies of the stellar populations (e.g. Chandar et al. 2002),
to X-ray surveys (e.g. Haberl \& Pietsch 2001).  However, in spite of M33
being the second closest spiral galaxy after M31, close enough for the
brighter members of its stellar population to be resolved, the properties of
its field halo stars have received comparatively little attention.

The first attempt to study the field halo stars in M33 using CCDs was
that of Mould \& Kristian (1986, hereafter MK86). In this classic paper, the 
authors observed
an approximately 5x5 arcmin field (see Fig. 1) with the Palomar 5m telescope
along with one of the first generation science-grade CCD detectors, a 
Texas Instruments 800 x 800 pixel array.  
Their color-magnitude diagram (CMD), based on aperture photometry of
215 stars 
transformed onto the Cousins system (Cousins 1976a,b), extends from
above the first ascent red giant branch (RGB) tip down to $\sim$1.5 mags
below the tip (I$\sim$22.5). From this diagram, they drew two primary
conclusions. First, by comparing the M33 RGB with those of the
Galactic globular clusters M92 and 47 Tuc, they found the mean metal
abundance of the field stars to be $\langle$[M/H]$\rangle$ = --2.2 $\pm$ 0.8.
This result was rather surprising given the much higher metallicity
($\langle$[M/H]$\rangle > -0.8$) they found for the halo of M31.
Second, from the bolometric magnitude of the RGB tip and an adopted
reddening of $E(V-I) = 0.06$, they
calculated an absolute distance modulus of 24.8 $\pm$ 0.2 for M33.

Cuillandre, Lequeux, \& Lionard (1999) used the UH8k CCD camera at
the prime focus of the CFHT 3.6m to image a 28 x 28 arcmin field in a region
of M33 that included the MK86 field. Their $(V,V-I)$ CMD reaches as faint 
as V $\sim$ 25.5. Cuillandre et al.
(1999) adopt an M33 distance modulus of $(m-M)_0 = 24.82$ and estimate
the line-of-sight reddening ($E(V-I) = 0.08$) by averaging values
based on the 21-cm line (Hartmann 1994) and from foreground stars 
(Johnson \& Joner 1987). Based on these values, a comparison with the
RGB sequences of Da Costa \& Armandroff (1990) yields a mean
metal abundance of [Fe/H] = --1.0 with a metallicity spread from
$\sim$--1.5 to $\sim$--0.6 dex for the field halo stars in M33.
This mean abundance is clearly much higher than the value advocated
by MK86 and a few tenths of a dex higher than the peak abundance of
Milky Way field halo stars (e.g. Norris 1994; Carney et al. 1996;).

Most recently, Davidge (2003) has used the Gemini North 8m telescope
equipped with GMOS to image a field located $\sim$9 kpc in projected
distance from M33 ($\sim$15.5 kpc deprojected distance), approximately 
twice as far out as the MK86 field. The g', r',
i', z' CMDs of Davidge (2003) are based on a field of view of only
5x5 arcmin and are not extraordinarily deep, reaching
some 2 magnitudes below the tip of the RGB, but they are notable as being
the first such M33 data at these extreme galactocentric distances.
Davidge's conclusions are twofold. 
First, based on the color of the upper RGB, Davidge (2003) determines 
a mean abundance of [Fe/H] = $-1.0 \pm 0.3$ (random) $\pm 0.3$ (systematic)
for this region of M33. Second, the presence of a significant number of
bright asymptotic giant branch stars suggests that an intermediate-age
population exists outside of the `young star-forming' disk of M33.
This latter result is consistent with the work of several investigators
(e.g. Sarajedini et al. 1998;2000, Chandar et al. 2002) who have suggested
that the halo star clusters in M33 are several Gyr younger than similar
clusters in the Milky Way's halo.

Within this context, the present series of papers attempts to shed further light
on the field halo stellar population(s) of M33. This first paper presents
deep ground-based photometry of the MK86 field along with an analysis
designed to probe the metallicity distribution function of this region of M33.
We begin with a discussion of
the observational material in the next section along with details of the
photometric reduction procedure. Section 3 describes our artificial star
experiments and how they are used to gauge the photometric 
completeness and errors. We compare our photometry to that of
MK86 in Sec. 4. The analysis portion of the paper begins in Sec. 5 in which
we use the luminosity function of the RGB to estimate the distance
to M33. In Sec. 6, the radial variations of the stellar populations present
in our field are analyzed. Section 7 presents the metallicity distribution
function and what it implies for this region of M33. Finally, our results
are summarized in Sec. 8.

\section{Observations, Image Reduction, and Photometry}

We obtained deep V and I images of a field centered about
$20\arcmin$ southeast of the center of M33, or about 10 kpc in deprojected
distance. The pointings were centered at $\alpha=01$:35:15, 
$\delta=+30$:30:00 J2000, chosen to sample the halo 
field population of M33 and
to overlap the photometry obtained by MK86 and so facilitate
comparison with their work.  Figure 1 shows the position of
our field (large square) and the position of the field studied in MK86
(small square) both relative to the center of M33 (upper right).  

Our data were obtained on the WIYN 3.5m
telescope 
at Kitt Peak National Observatory on the night of 1999 March 10 UT as part
of the `2-hour queue.'  The instrument was the S2KB CCD imager 
which provided a $0\farcs2$ pixel$^{-1}$ plate scale.  The CCD is a
$2048 \times 2048$ array and provided a $6\farcm8 \times 6\farcm8$ field.  
The seeing was generally excellent at $\sim0\farcs7$ allowing us to take full 
advantage of the fine plate scale.  A log of the observations is provided in 
Table 1. Figure 2 is a reproduction of one of our V-band CCD images.

The night of 1999 March 10 UT was not perfectly photometric, therefore
in order to photometrically calibrate the data, we obtained a set of
Landolt standards (Landolt 1992) and short exposures of the same M33
fields on 2002 February 11 UT.  These calibration data were taken with
the WIYN 0.9m telescope on Kitt Peak  using the same
S2KB CCD imager mounted at the Cassegrain focus.  The instrument 
configuration was identical but on the 0.9m telescope it has a plate
scale of $0\farcs6$ pixel$^{-1}$, and a $20\farcm5 \times 20\farcm5$
field of view.  Seeing was $\sim 1\farcs5$ but with 
the shallower exposures crowding was not a problem.  A log of these 
observations are also provided in Table 1.

We did all preliminary data reduction on both sets of images using the IRAF 
routines contained in {\it ccdproc}.  The raw CCD frames were processed to 
fit and remove the overscan level, trim the overscan region, subtract
a zero exposure time bias, and finally to flatten the individual pixel
responses with appropriate dome flats.  The final, flattened images were
all flat to $\lesssim0.2\%$ with the exception of the deep, 3.5 m $I$-band
images which had fringing at the $\sim 0.5\%$ level.  Since the
peak-to-peak pixel extent of the fringing was much greater than the typical
FWHM of point sources in the frame, we made no attempt to correct it.

We measured the aperture photometry of the SA98 standard stars using
the IRAF routine {\it phot}.  In each set of exposures of the SA98 field,
35 standards were measured so that the final calibration for the night
was determined from 70 measurements.  The night was determined to show
no evidence of a $(V-I)$ color term, to $\lesssim 0.5\%$, so we simply
fit the equations:
\begin{eqnarray}
v = V + v_1 + v_2 \times X_V \nonumber \\
i = I + i_1 + i_2 \times X_I \nonumber
\end{eqnarray}
where the capital letters denote standard magnitudes, the lower case
letters denote instrumental magnitudes, the $X$ variables are the
respective airmasses, and the numbered subscript coefficients are the fit 
coefficients.  The values determined for the calibration coefficients 
along with their uncertainties are given in Table 2.  From 
these uncertainties and the formal error aperture photometry measurements, 
the typical zero point uncertainty from these calibration equations 
is $\sim1\%$.

After photometrically calibrating the night of 2002 February 11 UT, we
calculated a photometric transformation between the shallow M33 exposures
taken on that night and the deep M33 exposures taken on the night of 1999
March 10 UT.  We measured the instrumental magnitudes for the deep exposures
using the ALLSTAR routine from DAOPHOT II (Stetson 1987).  A transformation 
for each deep exposure was calculated independently due to changing airmass, 
seeing, and possible transparency changes.  
There were typically $\sim 25$ stars in each transformation.  The coefficients 
of these transformations are given in Table 3.  Note that we found no 
color term in the $I$-band transformations so they are simple zeropoint 
offsets.  The $V$-band did have a $(V-I)$ color term of $\sim 3.5\%$.  
The $V$-band transformations were calculated by solving a least squares fit to
the equation:
\[
(V_{stand} - v_{inst}) = c_1 + c_2 \times (V-I)_{stand}
\]
and the $I$-band transformations were calculate by finding the mean
difference $\langle(I_{stand} - i_{inst})\rangle$.  To exclude saturated
stars in the deep 3.5m data, blends in either data set, mismatches between
the sets, and any stars contaminated by cosmic rays, the least squares fits 
for the $V$-band data were $2\sigma$ clipped and iterated until no additional 
stars were eliminated and the $I$-band mean differences were $2.5\sigma$ 
clipped.  The number of stars listed in the last column of Table~\ref{trans} 
are the number surviving the clipping and used to calculated the calibration.
The typical zero-point uncertainty in the calibration of the frames is 
$\sim 0.03$ mag.

After culling the data to remove spurious detections, extragalactic sources,
and unreliable detections, (see discussion below) the final calibrated data 
set contains 19,350 stars, measured in both $V$ and $I$.  Table 4
presents all of these data and Fig. 3 presents an $I$ versus $(V-I)$
color magnitude diagram.  The CMD shows a well defined upper giant branch
with a first ascent giant branch tip around $I\sim21$, and a population of 
asymptotic giant branch (AGB) stars above the tip. The upper giant 
branch shows inherent scatter larger than that expected from just photometric
error, and so there is likely to be an intrinsic dispersion in the age and/or
metallicity of the sample.  This is also apparent in the significant
number of blue main sequence stars (i.e. $I=23$-$24$ and $(V-I)\sim 0$),
the presence of stars brighter than the RGB tip, and the appearance of
a red clump dominated by core-helium burning stars at I$\sim$24.3.
In the sections below we examine these features in detail.

Although we do not possess a CMD of a field away from M33, we can simulate
the appearance of such a diagram using the Galaxy model of 
Robin et al. (1996, and references therein). Using the latest version of
their model available from their web page,\footnote{
$http://www.obs-besancon.fr/www/modele/modele\_ang.html$} we
have generated colors and magnitudes for stars in the line of sight
to our M33 field. These are shown as the gray points in Fig. 4 overplotted
on our M33 CMD from Fig. 3. Note that the simulated CMD includes
the effects of photometric error (Fig. 7) but not incompleteness (Fig. 6).
Figure 4 shows that some of the 
scatter seen redward of V--I $\sim$ 2 in the range 22$\lea$I$\lea$24
is due to field star contamination. In addition, although some stars
above the first ascent RGB tip likely do not belong to M33, the majority
of the AGB population with V--I$\gea$2.2 probably does. Lastly, we note 
that essentially all of the blue stars (V--I$\lea$0.4) are likely to belong to
M33.

\section{Completeness and Photometric Error Analysis}

In order to understand details about the CMD for
a field with such a complex mixture of stellar populations, we need
to quantitatively determine the photometric accuracy and star detection
efficiency of our photometric measurements.  To do this we created
a set of artificial stars of known magnitude and placed them on
each data frame and then measured their photometry using a procedure
identical to the procedure used with the real stars.  Since
image crowding may be a limiting factor, we placed the
artificial stars on the images in a regular grid pattern, so that no
where on the frames do they significantly increase the image crowding.

\subsection{Generation of Artificial Stars}

We generated the artificial stars to be introduced into the data frames
via the following procedure.  First we fit analytic functions to
three different regions of the observational CMD as displayed in
the left panel of Fig. 5.  Region one (R1) is the lower 
asymptotic giant branch which we fit with a second order polynomial 
in $I$.  Region two is the upper giant branch, which we fit with a line 
in $I$.  Region three is dominated by young main sequence stars.  The scatter 
proved too large for any kind of formal fit;
as a result, we represented it using a line from $23 \leq I \leq 24.5$.  In
both R1 and R2 the fits were least squares fits that were $2\sigma$ clipped
with no iterations.

We then used these analytic expressions to populate each section
of the CMD.  We extrapolated the R1 fit up to $I=20$ and distributed
540 stars randomly along the fit between $20 \leq I \leq 22$.  We only
used this relatively small number of stars because completeness is not
at question in this magnitude range.  We only needed to test the photometric 
accuracy.  We extrapolated the R2 fit down to $I=26$ to be sure to populate
the artificial stars to magnitudes well below our detection limit.  Since
this set is to be used for the determination of completeness as well
as photometric accuracy, we randomly distributed 9135 stars along the
extrapolated fit throughout this entire range.  The line segment representing
the stars in R3 was randomly populated with 325 stars.  In total, we
generated 10,000 artificial stars.

The 10,000 artificial stars were distributed across each frame in
a $100 \times 100$ grid.  The origin of this grid was pixel $(50,50)$
in the first $V$ frame with stars distributed every 19.5 pixels along
each axis.  The exact positions of the stars were their respective grid
center $\pm 0.5$ pixels randomly determined for each star so that the
positioning of each star relative to the pixel grid would not be regular.
Magnitude offsets between each frame were
determined and the coordinate transformation of each subsequent frame
onto the system of the first V frame (v1) was calculated; the
input coordinates and magnitudes of the artificial stars were then appropriately
offset.  We used the DAOPHOT II routine ADDSTAR to calculate the flux
and place each artificial star in each of the 7 data frames. 

We measured the photometry and processed the photometry lists from the
7 frames containing the artificial stars in a manner identical to the
original data frames.  The resulting
photometry of the artificial stars is shown by the black points in
the right panel of Fig. 5.  The solid grey lines in the
right panel indicated the input colors and magnitudes, so the scatter
in the plot gives a sense of the ultimate photometric accuracy, with
the caveat that the vast majority of the points are essentially
coincident with the grey lines.

\subsection{Completeness}

To get a complete understanding of our measured photometry,
we need to understand our detection efficiency as a function
of magnitude (i.e., the completeness).  The $I$ magnitudes of the artificial
data were generated randomly with emphasis on the dimmer part of the 
magnitude range, and then corresponding $V$ magnitudes were calculated to 
place each star along one of the sequences described above.  As a result 
the input $I$ magnitudes span $20 \leq I \leq 26$ whereas the $V$ magnitudes 
do not go as bright and only span $22.3 \lesssim V \lesssim 26.4$.  

Figure 6 is a plot of the completeness as both a function of $V$ 
and $I$ in 0.25 mag bins.  The $V$ detections are $100\%$ complete down to 
the 22.75--23 mag bin and the $I$ detections are $100\%$ complete down to 
the 21.75--22 mag bin.  Both bands then have a number of magnitude bins
where the completeness is very nearly $100\%$, and then fall off quickly in 
the usual fashion.  

All detections in each frame, real and artificial stars, were filtered
through a set of statistical parameters to prevent non-stellar
detections from appearing in our dataset.  In addition to reporting an instrumental
magnitude and error for each measurement, ALLSTAR also calculates 
a goodness of fit to the empirical PSF of the detection ($\chi$) and a 
measure of the similarity of the detection to a point source ($sharp$).
To avoid keeping measurements of bad pixels, cosmic ray hits,
error peaks in the halos of saturated stars, and extended objects
like background galaxies, we filtered the ALLSTAR output lists
to remove all objects with ${\rm err} > 0.6$ mag or ${\rm err} >
2\sigma \times \langle{\rm err}\rangle$ at a given magnitude, $\chi > 3$,
and $\vert {\rm sharpness}\vert > 0.5$.  Finally, both real and
artificial stars are kept in the dataset only if they appear on
at least 2 of the 3 $V$ images and 3 of the 4 I images.  (In the case
of the artificial stars, they were also considered found if they fell off
the edge of a given frame but were found in all of the frames they did
fall upon.)  As a result, the detection probability falls significantly 
below $100\%$ near $V \sim 25$ and $I \sim 24$. 

\subsection{Photometric Error}

A comparison of the absolute value of the difference between the input 
magnitudes and the measured magnitudes and the error estimates reported
by ALLSTAR shows that the two are in good agreement.  Figure 7
illustrates the comparison between the reported errors (lower panels), and the mean
of the absolute value of the difference between the input magnitudes
and the measured value averaged in bins 0.25 mag in width.  The error
bars in the top panels are the scatter in each bin.  The curves
are natural logarithm least-squares fits to the reported errors in
the lower panels, which are plotted over the binned measured errors
in the top panels.  Note that these curves are in good agreement with
the measured errors.  The only significant deviation is at magnitudes
dimmer than $I=25$ and $V=26$ both of which are beyond the point where
our completeness of detection begins to fall off rapidly.



\section{Comparison with the Photometry from Mould and Kristian}


To make a direct comparison of our photometric measurements to those
of MK86, we used their finder chart and data tables 
to find the coordinate transformation between the two data sets.
We then calculated the difference between our measured $V$ and $I$
photometry and theirs.  Excluding differences $> 1.0$ mag 
which are likely mismatches or extreme variable stars, the $I$
band comparison had 197 stars and the $V$ band 188.  We fit
equations of the form:
\[
(m_{\rm TSB} - m_{\rm MK}) = c_1 + c_2 \times (V-I)_{\rm TSB}
\]
where TSB denotes values from our photometry and MK denotes
values from Mould and Kristian.  The resulting fit coefficients
are tabulated in Table 5.  The $I$ band transformation
has a small but definite color term, but the $V$ band transformation
has a much less significant color term.  As an alternative, we also
calculated a simple zeropoint offset between the two data sets.
These values are also presented in Table 5.  The $I$ band
offset is essentially zero, but the scatter in the residuals is larger
than the scatter around the fit which includes the color term.  The
$V$ band has a 0.121 magnitude zero point offset, but in spite of its
color term being small in the linear fit, the scatter in the residuals
of the simple zeropoint offset are also larger than about the linear
fit.

Whichever transformation is used, the scatter in the residuals
is always $\gtrsim 0.2$ mag for both bands.  This suggests that the uncertainty
in Mould and Kristian's photometric measurements was a little larger than
they estimated, and likely a result of unresolved star
light contaminating their aperture photometry.  Additionally, the color term
in the I band transformation means that there is an offset between our
(V--I) colors and those of MK86.  Some of the conclusions we draw from 
our data are
different than those of MK86 primarily because of these disagreements
in the photometry.

\section{Luminosity Function and the Tip of the Giant Branch}

We constructed the $I$-band luminosity function of our field by
selecting stars from the $I$ versus $(V-I)$ CMD (Fig. 3).
We selected all stars brighter than $I=24$
that fell within a band 0.75 mags in width centered on the first
ascent giant branch locus.  We also included any stars brighter than
$I=21$ which fell redward of the blue edge of this band extrapolated
up to $I=18$.  The resulting luminosity function is displayed in
Fig. 8.  Based on the simulated field CMD shown in Fig. 4,
we estimate that the number of contaminating field stars in this
LF is small compared to the number of M33 stars.
This is supported by the findings of MK86,
who reached the same conclusion.


A number of features are apparent from this luminosity function.
The first ascent giant branch is approximated well by a power
law.  A power law fit to the bins in the range $20.75 \leq I \leq 24$
results in a line with slope, $\alpha = 0.431 \pm 0.008$, as indicated
by the heavy solid line in the figure.  We restricted the luminosity
function to magnitudes brigher than $I=24$ because this is our $>$95\%
completeness limit (see Fig. 6). Inspecting the
bins at brighter magnitudes, the first bin to significantly depart from
the power law is the bin centered at 20.625.  This suggests that the
tip of the first ascent giant branch (TRGB) must fall somewhere between
this bin and the 20.875 mag bin so that $I\sim20.75$.  
Stars brighter than
this tip are mostly AGB stars in M33 up until $I\sim 19.75$ at which
point the level reaches that expected for forground contamination.

The tip of the red giant branch is the last stage of a star's life
before it begins He burning on the zero-age horizontal branch.
Many studies have shown that $M_I^{\rm TRGB}$, the $I$-band absolute
magnitude of the TRGB, is remarkably constant over a wide range of
metallicities and ages (Da Costa \& Armandroff 1990;
Lee et al. 1993; Bellazzini et al. 2001).
Hence, one can estimate $I_{\rm TRGB}$, the $I$-band apparent
magnitude of the TRGB, to obtain a direct measurement of the
apparent distance modulus for a stellar population.

To locate $I_{\rm TRGB}$ in a systematic and reproducable way, 
we convolved the LF of stars in the color range
$\rm 1.2 < (V-I) < 3.0$ with a Gaussian-smoothed Sobel filter
weighted by the Poisson noise at each magnitude 
(Sakai, Madore, \& Freedman 1996; Sakai et al.\ 1997).  
The edge detector is very 
sensitive to noise in the LF so one must
restrict attention to a small range of magnitudes which 
are likely to include the TRGB.  We chose to examine the
range $20.75 \pm 0.4$ mag because it encompasses previous 
estimates (Mould \& Kristian 1986; Lee et al.\ 1993; Kim et al.\ 2002)
and based on our examination of the LF above.
Figure 9 shows the logarithmic LF and edge detector response in this 
magnitude range.  The highest peak in the filter 
response occurs at $\rm I_{\rm TRGB} = 20.75 \pm 0.02$.

To estimate this random error we ran a series of
bootstrap simulations in a manner similar to that of
$\rm M\acute{e}ndez$ et al.\ (2002).  The LF was randomly
resampled with replacement and each star was perturbed by 
randomly drawing from a Gaussian
distribution with standard deviation equal to the star's error.  
The edge detector was then applied to the resulting new LF.  
We repeated this process 50 times resulting in 50 estimates 
of the TRGB.  We found that these estimates were well fit 
by a Gaussian with a standard deviation of
0.02 mag which we adopt as the random error in our estimate 
of $I_{\rm TRGB}$.

Given this value of $I_{\rm TRGB}$, we now proceed to determine 
$M_I^{\rm TRGB}$ as follows. 
We use Eq. 4 of Bellazzini et al. (2001),
\[
M_I^{TRGB} = 0.14 {\rm[Fe/H]}^2 + 0.48 {\rm[Fe/H]} + 3.66,
\]
 to calculate $M_I^{\rm TRGB}$ using
an initial guess for the peak metallicity of the population. We then use the
resultant distance to construct a metallicity distribution function as described
in Sec. 7. The peak metal abundance of this distribution is input into the
above equation as the next guess for the peak [Fe/H]. This iterative procedure 
converges quickly to a solution yielding $M_I^{\rm TRGB} = -4.02 \pm 0.05$.  
Combined with our value for $I_{\rm TRGB}$, we find 
$(m - M)_I = 24.77 \pm 0.06$, where the quoted uncertainty includes errors 
in the measurement of $I_{\rm TRGB}$ along with random (0.02 mag) and 
systematic (0.03 mag) errors in the photometry. With our adopted reddening 
of $E(V-I)=0.06 \pm 0.02$ and $A_I = 1.31 E(V-I)$ (von Hippel \& Sarajedini
1998), we obtain 
an intrinsic distance modulus of $(m-M)_0 = 24.69 \pm 0.07$ for M33.
This value is within the range of distances tabulated by Kim et al. (2002).
However, it is significantly greater than the result of McConnachie et al.
(2004), who find $(m-M)_0 = 24.50 \pm 0.06$ also based on the TRGB.
The reasons for this difference are unclear at the moment.

\section{Radial Population Variations}

Adopting the position angle (23$^o$) and inclination (56$^o$) of M33 
derived by Regan \& Vogel (1994) and the distance modulus based on
the TRGB, we have computed deprojected radial positions 
relative to the center of M33 for each star in our sample. Using these, 
Fig. 10 shows CMDs for our M33 data divided into four radial bins. The
dashed boxes represent the main sequence (MS) region (19.0$\leq$I$\leq$23.5, 
--0.5$\leq$(V--I)$\leq$0.5), the asymptotic giant branch (AGB) region
(19.7$\leq$I$\leq$20.7, 1.4$\leq$(V--I)$\leq$3.5), and the red giant branch
(RGB) region (20.85$\leq$I$\leq$22.85, 1.0$\leq$(V--I)$\leq$2.15). The MS is
representative of stars with ages younger than $\sim$0.5 Gyr, the AGB is mainly
stars between $\sim$2 Gyr and $\sim$8 Gyr, and the RGB is dominated by 
stars older than $\sim$8 Gyr (see Fig. 12). Inspection of Fig. 10 suggests
that the mean age of the M33 stellar population is increasing as we
proceed to larger galactocentric distances in our observed M33 field. 

In order to quantify this impression, Fig. 11 shows the radial behavior
of the various `age tracer' populations we have identified above. The filled 
squares in Fig. 11 show the variation in the stellar density of
all stars that are above our 95\% completeness limit (V$\lea$25,
I$\lea$24). The open circles, open squares, and filled circles represent
the RGB, AGB, and MS regions, respectively. For each region, we have 
subtracted off the foreground field-star contamination determined 
from Fig. 4 and then divided by the 
total stellar density at each radius to account for the overall decrease in the 
density as R increases. The lines are the weighted least squares fits
to each set of points. We see that the inner regions of our M33 field are,
in the mean, younger than the outer regions; this is most dramatically
illustrated by examining the MS region.

To estimate the age gradient in our field, we proceed as follows. First, we note
that the age gradients implied by the three CMD regions (MS, AGB, RGB)
must be consistent with each other. As such, we concentrate on the MS region
because the gradient is most conspicuous there. Based on the locations of the 
isochrones in the MS region of the CMD in Fig. 12, we estimate an upper limit 
of 1 Gyr for the total age gradient in our M33 field. We will investigate this
question further in Paper II where we present a synthetic CMD analysis we 
have performed using the starFISH code of Harris \& Zaritsky (2001).

\section{Metallicity Distribution Function}

It is instructive to examine the distribution of RGB stars as compared with
empirical RGB sequences for a set of standard Galactic globular clusters (GC). 
This approach has historically been used to derive a metallicity distribution
function (MDF) for stellar populations (e.g. Durrell, Harris, \& Pritchet 2001;
Sarajedini \& Van Duyne 2001; Harris \& Harris 2002). It involves using
the dereddened color of RGB stars as a surrogate for metal abundance
under the assumption that all stars on the dominant RGB sequence
are old (i.e. $\gea$10 Gyr). In this section, we apply this technique
to our CMD and investigate the properties of the resultant MDF.

\subsection{Construction and Global Properties}

The upper panel of Fig. 12 shows our
M33 CMD along with Z=0.004 ([Fe/H]$\sim$--0.7)
theoretical isochrones (Girardi et al. 2002) for
ages of $10^7$, $10^8$, $10^{8.5}$, $10^9$, and $10^{10}$ years. A distance 
modulus of $(m-M)_I = 24.77$ and a reddening of $E(V-I) = 0.06$ have
been applied to the photometry. We see that stars
with a range of ages are present in our field. However, the RGB stars appear
to be dominated by a population that is of order $10^{10}$ years old, which
means that the majority of the resultant stellar metallicities will not be
significantly affected by age. This assertion is supported by preliminary 
StarFISH synthetic CMD work we have performed. The full starFISH 
analysis and a complete 
discussion of the results will be presented in Paper II of this series. 

The lower panel of Fig. 12 illustrates our technique for constructing the 
MDF by showing empirical 
GC RGBs formulated by Saviane et al. (2000) for $[Fe/H]$ values of --2.0, 
--1.5, --1.0, and --0.7 dex. For each star with $-3.9 \geq$$M_I$$\geq -2.4$ and 
1.0$\leq$$(V-I)_0$$\leq$2.2, we can convert dereddened color to metal abundance 
using this grid of RGBs. 
We have chosen this range of color and magnitude to 
minimize the influence of old AGB stars that are bluer than the first 
ascent RGB. 
The result of this exercise is shown in Fig. 13. The filled circles in
the upper panel represent the binned histogram of metallicities for the 1123 RGB
stars that meet our selection criteria. The solid line is the generalized 
histogram of these values constructed by adding up unit Gaussians with widths
given by the error in each metalliciity. The generalized MDF has at least two
advantages over the binned one. First, it is not subject
to the vagaries associated with binning - the choice of endpoints and the 
bin sizes. Second, genuine (and artifactual) features of the distribution 
are revealed when the errors are taken into account in its construction.
We see that the MDF exhibits a prominent peak at [Fe/H]$\sim -1.1$, a 
possible secondary peak at [Fe/H]$\sim -1.4$, and a gradual tail to lower
metallicities.

To demonstrate the effects of reddening uncertainties on the construction of
the MDF, the lower panel of Fig. 13 shows the generalized MDF from the upper
panel (solid line) along with two additional distributions - one constructed 
using a reddening that is 0.02 mag smaller than our adopted value of 
$E(V-I)=0.06 \pm 0.02$ (dashed) and the other using a value that is 0.02 mag 
larger (dotted). We note that the essential features of the distribution 
discussed above are unchanged.

As noted above, our preliminary StarFISH results 
suggest that the ages of the RGB stars are not likely to significantly affect
the derived MDF. We estimate that the metallicity of the peak {\it could} be
biased downward by 0.1 to 0.2 dex due to the presence of 
stars younger than $\sim$8 Gyr in the RGB sample that produced the 
MDF. Thus, the true peak of the MDF could be as high as [Fe/H]$\sim -0.9$.
This number is somewhat uncertain, but it does
represent a likely upper limit on the effect of age on the MDF peak.

\subsection{Radial Abundance Variation}

Figure 14a plots the metal abundance of
each RGB star as a function of its deprojected radial location in kiloparsecs.
A robust
iterative 2-$\sigma$-rejection least-squares fit (Sarajedini \& Norris 1994) 
to these data reveals a modest, yet statistically significant, gradient yielding 
\[
[Fe/H] = -0.06 (\pm 0.01) R_{deproj} - 0.45 (\pm 0.08),
\]
as shown by the solid line in Fig. 14a. 

Figure 14b illustrates radial abundance gradients derived
from other M33 populations compared with our data. First, we divided
our data from Fig. 14a into two bins at R=10.5 kpc. We then fit the dominant
peak of the MDF in each bin with a Gaussian function. The resultant
peak metallicities are plotted as open circles in Fig. 14b.
The filled circles in Fig. 14b are the M33 disk regions studied by
Kim et al. (2002, hereafter KKLSG). They constructed CMDs based on 
Hubble Space
Telescope Wide Field Planetary Camera 2 observations of 10 M33 disk regions.
Their CMDs extend to I$\sim$27 and sample the disk stars in M33.
KKLSG followed a procedure similar to our's (thus allowing an intercomparison of
the results) and used the dereddened 
color of the RGB to estimate the mean metal abundance in each of their 
fields. A least-squares fit to all of the filled circles yields
\[
[Fe/H] = -0.05 (\pm 0.01) R_{deproj} - 0.55 (\pm 0.02)
\]
shown by the solid line. If we exclude the two innermost points, where the image
crowding may have been problematic, we find
\[
[Fe/H] = -0.07 (\pm 0.01) R_{deproj} - 0.48 (\pm 0.04),
\]
which is illustrated by the long dashed line. In order to reliably compare these
abundance gradient relations with our result, we need to adjust our
data for the difference in distance modulus between the present
work [$(m-M)_I = 24.77$] and that of KKLSG [$(m-M)_I = 24.89$]. 
Having done this, we find that 
\[
[Fe/H] = -0.07 (\pm 0.01) R_{deproj} - 0.49 (\pm 0.09),
\]
which is statistically identical to both of the lines derived by KKLSG.
This suggests that the vast majority of stars in our M33 field belong to
the disk population of that galaxy, {\it not} the halo as previously
thought (e.g. MK86). If this assertion is borne out, then the M33 stellar disk
would extend out to $\sim$12 kpc.

At this point, it is important to note that the slight radial age variation
determined in Sec. 6 would have a negligible effect on our metallicity gradient 
(Fig. 14a). The Girardi et al. (2002) isochrones predict a change in the
implied metallicity of $\sim$0.05 dex/Gyr due to the effect of age on the
colors of RGB stars at $M_I$=--3.5.
Given our upper limit of 1 Gyr for the total age gradient in our field, this implies
a differential correction of $\pm$0.025 dex from the inner regions
to the outer regions. The resultant ``corrected" slope would be 
$\Delta$[Fe/H]/$\Delta$$R_{deproj}$ = --0.07, which is still consistent with
the equations given above.

For reference, the dotted line in Fig. 14b shows the radial abundance 
variation of the HII regions as derived by Kim et al. (2002, see their Fig. 5). 
Lastly, the open squares in Fig. 14b are 
the nine M33 halo globular clusters from Sarajedini et al. (2000). They exhibit
no radial abundance gradient. 
This is one signature of a chaotic fragmentation type formation scenario similar
to Galactic globular clusters outside 8 kpc from the Galactic center, which also
show no radial abundance gradient (Searle \& Zinn 1978). As pointed out by 
Sarajedini
et al. (1998,2000), the M33 halo clusters suffer from the second parameter 
effect again analogous to the outer Galactic halo globulars. If our M33 field
was dominated by halo stars, then there would be no significant radial
abundance gradient among the stars just like the halo clusters. 

\section{Summary and Conclusions}

In this work, we present deep $VI$ photometry of a field in M33,
that includes the region studied by Mould \& Kristian(1986), observed with the 
WIYN 3.5m telescope. Based on the CMD, which reaches past
the helium burning red clump, we draw the following conclusions.

\noindent 1) The I-band apparent magnitude of the TRGB is measured to
be $\rm I_{\rm TRGB} = 20.75 \pm 0.04$, where the error includes the
measurement error (0.02 mag) along with random (0.02 mag) and 
systematic (0.03 mag) errors in the photometry. 

\noindent 2) The I-band apparent magnitude of the TRGB
implies  a distance modulus of $(m-M)_I = 24.77 \pm 0.06$, which combined
with an adopted reddening of $E(V-I)=0.06\pm0.02$ yields an absolute modulus
of $(m-M)_0 = 24.69 \pm 0.07$ (867$\pm$28 kpc) for M33.

\noindent 3) By examining the spatial properties of stars in 
various parts of the CMD, we find that, over the range
of deprojected radii covered by our field ($\sim$8.5 to $\sim$12.5 kpc),
there is a significant age gradient with an upper limit of 
$\sim$1 Gyr (i.e., $\sim$0.25 Gyr/kpc). 

\noindent 4) Using the empirical RGB grid constructed by Saviane et al.
(2000), we have converted the dereddened color of each RGB star
to a metallicity. The resultant metallicity distribution function (MDF) 
displays a primary peak at a metallicity of $[Fe/H]$$\sim$--1.0 with a tail to
lower abundances. The peak shows a radial variation with a slope
of $\Delta$[Fe/H]/$\Delta$$R_{deproj}$ = --0.06 $\pm$ 0.01 dex/kpc.
This gradient is consistent with the variation seen in the inner {\it disk} regions 
of M33. Therefore, we conclude that the vast majority of stars in this field belong
to the disk of M33, not the halo as previously thought.

\acknowledgments

The authors wish to thank Constantine Deliyannis for taking the 
2002 February 11 UT photometric calibration frames. This research
was funded by NSF CAREER grant AST-0094048 to A.S.

\clearpage

\begin{figure}
\caption{A finder chart for our field (large box) and the relative positions 
of the Mould \& Kristian (1986) field (small box) and the center of M33.  North is up
and East is to the left.  The image is approximately $30\arcmin$ wide.
\label{finder}} 
\end{figure}

\clearpage

\begin{figure}
\epsscale{0.9}
\caption{A V-band CCD image of our M33 field. The field is 6.8 x 6.8 arcmin.
North is up and east to the left.}
\end{figure}

\clearpage

\begin{figure}
\caption{$I$ versus $(V-I)$ color magnitude diagram of the entire sample.
\label{cmd}}
\end{figure}

\begin{figure}
\caption{Same as Fig. 3 except that the simulated distribution of non-M33 stars
is overplotted as the filled circles.}
\end{figure}

\clearpage

\begin{figure}
\caption{Generation of artificial stars.  The left panel shows the three
regions in which analytic functions were fit to the observed distribution
of stars.  Region 1 (R1) contains the AGB.  We fit a second order polynomial 
using $2\sigma$ clipped least squares and then extrapolated the polynomial to
$I=20$.  Region 2 (R2) contains the upper giant branch.  We fit a line using
$2\sigma$ clipped least squares.  The line was then extrapolated down to
$I=26$.  Region 3 (R3) contains the young main sequence.  The scatter precluded
fitting any analytic function so a line bisecting the region was chosen.
The input data generated from the analytic functions are shown in the
right panel by the grey solid lines.  The black points indicate the 
measured artificial data.  See text for details.   \label{cmdart}}
\end{figure}

\clearpage

\begin{figure}
\epsscale{1}
\caption{Completeness as a function of magnitude.  The open circles
show the completeness of the $V$ band data and the filled circles
the $I$ band data, both in bins of 0.25 mag.  The $V$ detections are $100\%$ 
complete down to the 22.75--23 mag bin and the $I$ detections are $100\%$ 
complete down to the 21.75--22 mag bin.  Both bands then have a number 
of magnitude bins where the completeness is very nearly $100\%$, and
then fall off quickly in the usual fashion.  This protracted fall off 
is caused by the filtering process we preformed on all of the data to 
excluded erroneous detections.  See the text for details.  \label{comp}}
\end{figure}

\clearpage

\begin{figure}
\caption{Measured error and estimated error.  The bottom panels show
the photometric measurement errors reported by ALLSTAR for the artificial
data.  The curve in each plot is the least squares fit to the equation 
$\ln({\rm error}) = a + b\times{\rm mag}$.  These curves are plotted in
the respective upper panels which also show the mean difference between
the input magnitudes and the measured magnitudes binned in 0.25 mag 
bins.  The error bars are the scatter in each bin.  Note that
the estimated errors are a very good estimated of the actual errors
in measurment down to $I=25$ and $V=26$ at which point the measured magnitudes
are systematically brighter than the error estimates suggest.  These magnitudes
are much fainter than the level where the data are $100\%$ complete and so
well below the magnitude range we use in our analysis.  \label{perr3}}
\end{figure}

\clearpage

\begin{figure}
\caption{The $I$-band luminosity function for giant stars in M33.
The histogram is the $I$-band luminosity function for selected stars
from our $I$ versus $(V-I)$ CMD (see text for selection details).
The heavy solid line is a power law fit in the range $20.75 \leq I \leq 24$,
and has a slope, $\alpha = 0.431 \pm 0.008$.  The first significant
departure from this power law toward brighter magnitudes occurs between
the bins centered at 20.625 and 20.875 suggesting that the tip of the
first ascent giant branch is around $I=20.75$. \label{lfsel}}
\end{figure}

\clearpage

\begin{figure}
\caption{A plot of the results from the edge detection algorithm.  In the top
panel is the logarithm of the Gaussian-smoothed luminosity function for 
color-cut stars in the neighborhood of the TRGB.  The bottom panel shows the
edge detector response.  The vertical line marks the position of the TRGB 
at $I = 20.75$ (see text for details). \label{edge}}
\end{figure}

\clearpage

\begin{figure}
\epsscale{0.95}
\caption{The radial variation of the M33 CMD in our field. The data are
arbitrarily divided into four radial bins based on deprojected radial distances
in kiloparsecs. The dashed boxes represent the
main sequence (MS) region (19.0$\leq$I$\leq$23.5, 
--0.5$\leq$(V--I)$\leq$0.5), the asymptotic giant branch (AGB) region
(19.7$\leq$I$\leq$20.7, 1.4$\leq$(V--I)$\leq$3.5), and the red giant branch
(RGB) region (20.85$\leq$I$\leq$22.85, 1.0$\leq$(V--I)$\leq$2.15). The MS is
representative of stars with ages younger than $\sim$0.5 Gyr, the AGB is mainly
stars between $\sim$2 Gyr and $\sim$8 Gyr, and the RGB is dominated by 
stars older than $\sim$8 Gyr (see Fig. 12).}
\end{figure}

\clearpage

\begin{figure}
\epsscale{1}
\caption{The radial variation of star counts in our M33 field. The filled
squares in the upper portion of the diagram show the behavior of all 
stars above our 95\% completeness limit (V$\leq$25, I$\leq$24). The stellar
densities in the MS (filled circles), AGB (open squares), and RGB (open
circles) regions are scaled by the total number at each radius and plotted in the
lower part of the figure. The contribution from stars in the line of sight
has been subtracted off at each radius. The lines represent weighted least 
squares fits to each set of points.}
\end{figure}

\clearpage

\begin{figure}
\epsscale{1}
\caption{The upper panel shows our M33 CMD along with the
Girardi et al. (2002) isochrones for Z=0.004 and ages of 
$10^7$, $10^8$, $10^{8.5}$, $10^9$, and $10^{10}$ years.
The lower panel illustrates our M33 CMD along with the Saviane et al. 
(2000) empirical 
RGB grid for $[Fe/H]$ values of --2.0, --1.5, --1.0, and --0.7 dex.}
\end{figure}

\clearpage

\begin{figure}
\epsscale{1}
\caption{The upper panel shows the metallicity distribution function (MDF) for our
M33 field. The filled circles represent the binned histogram while the solid
line is the generalized histogram. The MDF has been constructed by
comparing RGB stars in the range $-3.9 \geq$$M_I$$\geq -2.4$ and 
1.0$\leq$$(V-I)_0$$\leq$2.2 with the Saviane et al. (2000) empirical RGB grid.
The lower panel illustrates the effects of changing the adopted reddening
by $\pm$0.02 mag in E(V--I).}
\end{figure}

\begin{figure}
\epsscale{0.85}
\caption{Plots of metal abundance as a function of deprojected
radial position in kiloparsecs for various populations in M33. The
open circles in the upper panel represent individual RGB
stars in our M33 field. The solid line is an iterative 2-$\sigma$ rejection 
least-squares fit to these data showing a significant abundance 
gradient in this range of galactocentric distances. The filled
circles in the lower panel are the M33 disk regions studied by
Kim et al. (2002). The solid line is the fit to all of
the filled circles while the dashed line is the fit excluding the inner two
points. The dotted line is the radial abundance variation of the
HII regions as derived by Kim et al. (2002). The open squares are 
the nine M33 halo globular clusters from Sarajedini et al. (2000). 
Lastly, the open circles in the lower panel are mean metallicities
and radial positions of the M33 stars shown in the upper panel.}
\end{figure}

%


\clearpage

\begin{deluxetable}{llcrrr}
\tablecaption{Log of Observations\label{obs}}
\tablewidth{0pt}
\tablehead{
   \colhead{Field}
  &\colhead{Date}
  &\colhead{Band}
  &\colhead{Exposure}
  &\multicolumn{2}{c}{Field Center (J2000)}
  \\[0.0ex]

\colhead{}
  &\colhead{(UT)}
  &\colhead{}
  &\colhead{Times (s)}
  &\colhead{RA}
  &\colhead{DEC}
}
\startdata
M33 & 1999-03-10 & V &  800 & 01:35:15 & 30:30:00 \\
M33 &            & V &  800 & 01:35:15 & 30:30:12 \\
M33 &            & V &  800 & 01:35:14 & 30:30:23 \\
M33 &            & I & 1100 & 01:35:14 & 30:30:23 \\
M33 &            & I & 1100 & 01:35:15 & 30:30:12 \\
M33 &            & I & 1100 & 01:35:15 & 30:30:00 \\
M33 &            & I & 1100 & 01:35:14 & 30:30:12 \\
SA98 & 2002-02-11 & V & 60 & 06:52:13 & -00:19:02 \\
SA98 &            & V & 60 & 06:52:12 & -00:19:16 \\
SA98 &            & I & 50 & 06:52:13 & -00:19:02 \\
SA98 &            & I & 50 & 06:52:12 & -00:19:16 \\
M33  & 2002-02-11 & V & 600 & 01:36:00 & 30:27:51 \\
M33  &            & V & 600 & 01:35:57 & 30:26:20 \\
M33  &            & V & 600 & 01:36:07 & 30:28:20 \\
M33  &            & I & 300 & 01:35:58 & 30:26:20 \\
M33  &            & I & 300 & 01:36:02 & 30:27:20 \\
M33  &            & I & 300 & 01:36:06 & 30:28:20 \\
\enddata
\end{deluxetable}

\clearpage

\begin{deluxetable}{lrr}
\tablecaption{Photometric Coefficients for 2002 February 11\label{photcoef}}
\tablewidth{0pt}
\tablehead{
   \colhead{Coefficient}
  &\colhead{Value}
  &\colhead{Uncertainty}
}
\startdata
v1 & 3.2913 & 0.0072 \\
v2 & 0.1337 & 0.0054 \\
i1 & 4.0270 & 0.0098 \\
i2 & 0.0826 & 0.0075 \\
\enddata
\end{deluxetable}

\clearpage

\begin{deluxetable}{lccc}
\tablecaption{Photometric Transformation Coefficients\label{trans}}
\tablewidth{0pt}
\tablehead{
   \colhead{Frame}
  &\colhead{Zero Point}
  &\colhead{Color Term}
  &\colhead{Number}
}
\startdata
V1 & $7.036 \pm 0.010$ & $0.040 \pm 0.007$ & 26 \\
V2 & $7.040 \pm 0.012$ & $0.033 \pm 0.008$ & 22 \\
V3 & $6.998 \pm 0.012$ & $0.039 \pm 0.008$ & 27 \\
I1 & $6.965 \pm 0.032$ & \nodata & 24 \\
I2 & $6.980 \pm 0.032$ & \nodata & 26 \\
I3 & $6.959 \pm 0.030$ & \nodata & 29 \\
I4 & $7.033 \pm 0.028$ & \nodata & 24 \\
\enddata
\end{deluxetable}

\clearpage

\begin{deluxetable}{rllrrrrr}
\tablecaption{Calibrated Stellar Photometry and Positions\label{data}}
\tablewidth{0pt}
\tablehead {
   \colhead{ID}
  &\colhead{RA}
  &\colhead{DEC}
  &\colhead{I}
  &\colhead{err}
  &\colhead{V-I}
  &\colhead{err}
  \\[0.0ex]
\colhead{}
  &\multicolumn{2}{c}{(J2000.0)}
  &\colhead{(mag)}
  &\colhead{(mag)}
  &\colhead{(mag)}
  &\colhead{(mag)}
}
\startdata
     1 &  01:35:02.47 &   30:31:14.8 &  16.587 &  0.005 &  0.594 &  0.006 \\
     2 &  01:35:15.53 &   30:29:39.2 &  16.739 &  0.012 &  0.903 &  0.013 \\
     3 &  01:35:25.45 &   30:32:28.8 &  16.838 &  0.016 &  2.118 &  0.018 \\
     4 &  01:35:03.77 &   30:30:04.7 &  17.107 &  0.008 &  1.888 &  0.012 \\
     5 &  01:35:14.13 &   30:31:50.6 &  17.298 &  0.005 &  2.489 &  0.007 \\
\enddata
\tablecomments{The complete version of this table containing all 19350
stars is in the electronic edition of the Journal.  The printed edition 
contains only a sample.}
\end{deluxetable}

\clearpage

\begin{deluxetable}{lcccc}
\tablecaption{Transformation Coefficients from Mould \& Kristian Photometry \label{mktrans}}
\tablewidth{0pt}
\tablehead{
   \colhead{Band}
   &{Zero Point}
   &{Color Term}
   &{Number}
   &{$1\sigma$}
   \\[0.0ex]
\colhead{}
   &{$(c_1)$}
   &{$(c_2)$}
   &{}
   &{resid.}
}
\startdata
\sidehead{Linear fits}
I band  & $0.127 \pm 0.039$ & $-0.065 \pm 0.019$ & 197 & 0.191 \\
V band  & $0.157 \pm 0.034$ & $-0.026 \pm 0.021$ & 188 & 0.232 \\
\sidehead{Weighted mean offsets}
I band  & $0.005 \pm 0.003$ & \nodata & 197 & 0.254 \\
V band  & $0.121 \pm 0.005$ & \nodata & 188 & 0.287 \\
\enddata
\end{deluxetable}

\end{document}